\documentstyle[aps,graphicx]{revtex}\tighten

\begin{document}
\title{Gravitational entropy in cosmological models}

\author{Nicos Pelavas\dag, Alan Coley\dag}
\address{\dag\ Department of Mathematics and Statistics, Dalhousie University, Halifax, Nova Scotia}
\address{email: pelavas@mathstat.dal.ca, aac@mathstat.dal.ca}

\maketitle

\begin{abstract}
We discuss whether an appropriately defined dimensionless scalar function might be an acceptable candidate for the
gravitational entropy, by explicitly considering Szekeres and Bianchi type VI$_{h}$ models that admit an isotropic
singularity. We also briefly discuss other possible gravitational entropy functions, including an appropriate measure
of the velocity dependent Bel-Robinson tensor.
\end{abstract}

\section{Introduction}
Penrose \cite{penrose} has argued that the initial cosmological singularity must be one of low entropy in order to
explain the high isotropy of the observed universe and to be consistent with the second law of thermodynamics. Since
the matter was presumably in thermal equilibrium at the initial singularity, this implies low entropy in the
gravitational field. Penrose has also conjectured that such a gravitational entropy should be related to a suitable
measure of the Weyl curvature. The search for a suitable candidate for the "gravitational entropy" is therefore of
current interest, particularly in the approach to the initial cosmological singularity \cite{tod,GCW}. In the quiescent
cosmological paradigm \cite{barrow} the universe began in a highly regular state and subsequently evolved towards
irregularity.

Penrose \cite{penrose} originally proposed that the the Weyl tensor is zero at the big bang singularity, implying the
subsequent evolution is close to a Friedman-Robertson-Walker (FRW) model. However, this requirement is too strong.  For
example, in perfect fluid spacetimes Anguige and Tod \cite{Anguige} have proven uniqueness results that show that if
the Weyl tensor is zero at the big bang, then the spacetime geometry must be exactly FRW in a neighbourhood of the
big-bang. This has motivated the idea that some appropriate dimensionless scalar is asymptotically zero. The search for
a gravitational entropy then  reduces to a search for this scalar function. Quiescent cosmology and the ideas of
Penrose provide the motivation of the definition of an isotropic singularity \cite{GW}; essentially a spacetime admits
an isotropic singularity if the 'physical' spacetime is conformally related to an 'unphysical' spacetime such that
there exists a time function $T$ with the property that at $T=0$ the conformal factor vanishes (corresponding to the
cosmological singularity) but that the conformally related metric is regular. It was proven in \cite{GW} that in the
class of models with an isotropic singularity $P \rightarrow 0$ as $T \rightarrow 0$, where $P$ is the ratio of the
Weyl curvature squared to the Ricci curvature squared  (see Eqn. (\ref{szekP}) below).

Recently, Lake and Pelavas (LP) \cite{hvfinv} considered a class of "gravitational epoch" functions that are a
dimensionless scalar field constructed from the Riemann tensor and its covariant derivatives only (such as, for
example, $P$). They discussed whether such functions can act as a "gravitational entropy" by determining whether it is
monotone along a suitable set of (smooth) timelike trajectories. In particular, LP considered the set of homothetic
trajectories of a self-similar spacetime (since such spacetimes are believed to play an important role in describing
the asymptotic properties of more general models \cite{Carrandcoley}). They showed that the Lie derivative of  any
"dimensionless" scalar along a homothetic vector field (HVF) is zero, and concluded that such functions are not
acceptable candidates for the gravitational entropy. They suggested considering other options for a "gravitational
epoch" function.  Other dimensionless scalars constructed from the Riemann tensor and its covariant derivatives have
been considered (see, for example, \cite{GCW}). Other alternatives, such as those involving the Bel-Robinson tensor,
were suggested in \cite{grqcbelrob}.

In this paper, by explicitly considering classes of Szekeres and Bianchi VI$_{h}$ models that admit an isotropic
singularity, we revisit the conclusion of LP that $P$ (for example) is not an acceptable candidate for the
gravitational entropy. First, we take the view, unlike that taken in LP, that a purely gravitational entropy selects
spacetimes as being of cosmological interest according to a thermodynamic principle. For example, in General relativity
(GR) there exist solutions for which the physical energy density is negative; however, we do not disregard GR or the
notion of energy density within GR -- rather, we use the criterion of negative energy density to characterize those
solutions which are not physical. Second, homothetically self-similar spacetimes represent asymptotic equilibrium
states (since they describe the asymptotic properties of more general models \cite{Carrandcoley}), and the LP result is
perhaps consistent with this interpretation since the entropy does not change in these equilibrium models, and perhaps
consequently supports the idea that $P$ (for example) represents a "gravitational entropy". Therefore, we will
investigate the behavior of  $P$ asymptotically as the self-similar solution is approached in cosmological models. We
also consider various other options for a purely ``gravitational entropy", including an appropriate measure of the
Bel-Robinson tensor.

\section{Szekeres model}
We consider the class II Szekeres solutions, which are spatially inhomogeneous models with irrotational dust as a
source. A comprehensive analysis of the Szekeres models can be found in \cite{gw82a}. In the notation of Goode and
Wainwright we set $k=0=\beta_{-}$, the line-element in comoving coordinates has the form

\begin{equation}
ds^2=T^4[-dT^2+dx^2+dy^2+(A-\beta_{+}T^2)^{2}dz^2]   \label{szek}
\end{equation}

\noindent where

\begin{equation}
A=a(z)+b(z)x+c(z)y-5\beta_{+}(z)(x^2+y^2).
\end{equation}

\noindent The fluid 4-velocity is $u=T^{-2}\frac{\partial}{\partial T}$ and the energy density is
$\mu=12T^{-6}[1-(\beta_{+}/A)T^2]^{-1}$. These cosmological models admit an isotropic singularity at $T=0$; moreover,
it has been shown \cite{gw82b} that the general Szekeres class with $\beta_{-}=0$ also admits an isotropic singularity.
If $\beta_{+}=0$ in (\ref{szek}), then we obtain the associated flat FRW dust solution.

The standard gravitational epoch function, $P$, for (\ref{szek}) is

\begin{equation}
P \equiv \frac{C_{abcd}C^{abcd}}{R_{ab}R^{ab}} = \frac{4}{3}\frac{T^{4}\beta_{+}^{2}}{A^2}.  \label{szekP}
\end{equation}

\noindent This has been shown \cite{gcw} to behave appropriately in these models, i.e. $P$ is monotonically increasing
away from the isotropic singularity. Perhaps an alternative choice is to use the Bel-Robinson tensor \cite{grqcbelrob}
to construct a velocity dependent gravitational epoch function. Using the fluid 4-velocity, we construct the positive
scalar

\begin{equation}
W = T_{abcd}u^{a}u^{b}u^{c}u^{d} = \frac{24\beta_{+}^{2}}{T^8(\beta_{+}T^2-A)^2}
\end{equation}

\noindent which has the same units as $C_{abcd}C^{abcd}$, and hence to obtain a dimensionless scalar we normalize by
the square of $\mu=T_{ab}u^{a}u^{b}$, i.e.

\begin{equation}
\widetilde{P} = \frac{W}{\mu^2}=\frac{T^4\beta_{+}^{2}}{6A^2}.
\end{equation}

Noting that in this case we have $\widetilde{P}=P/8$ then $\widetilde{P}$ also behaves appropriately as the isotropic
singularity is approached. This relationship is a consequence of the following two facts. First, the magnetic part of
Weyl, $H_{ab}=0$, therefore $C_{abcd}C^{abcd}$ is equivalent to $W$ modulo a positive constant.  Second, the Ricci
invariant $R_{ab}R^{ab}=R^2$ and for dust $R=\mu$.  In these models, the choices are limited for constructing
dimensionless ratios of zeroth order invariants since all Carminati-McLenaghan (CM) invariants \footnote{We note that
in this case any complex CM invariants have vanishing imaginary part.} can be expressed in terms of the Ricci scalar
$R$ and $Re(w_{1})\sim C_{abcd}C^{abcd}$,

\begin{eqnarray}
R=\frac{12A}{T^6(A-\beta_{+}T^2)} &,\ \ & w_{1} = \frac{24 \beta_{+}^{2}}{T^8(A-\beta_{+} T^{2})^{2}} \\
\end{eqnarray}

\noindent with syzygies for the Ricci invariants,

\begin{equation}
\begin{array}{ccccc}
r_{1}=\frac{3}{16}R^{2},  & & r_{2}=\frac{3}{64}R^3, & & r_{3}=\frac{21}{1024}R^4. \\
\end{array}
\end{equation}

\noindent Since these spacetimes are Petrov type D, the Weyl invariants satisfy $6w_{2}^2=w_{1}^3$.  The mixed
invariants give syzygies

\begin{equation}
\begin{array}{ccccccc}
m_{1}=0, & & m_{2}=\frac{1}{16}w_{1}R^{2}=m_{3}, & & m_{4}=-\frac{1}{128}w_{1}R^3, & & m_{5}=\frac{1}{16}w_{2}R^{2}. \\
\end{array}
\end{equation}

To study the Lie derivative of $P$ (or $\widetilde{P}$) along a HVF we first recall a well known result
\cite{Carrandcoley}.  Any FRW model with an equation of state $p=-\mu /3$ admits a timelike HVF.  In addition, only the
flat FRW models with an equation of state $p=\alpha \mu$ and power law dependence on the scale function admit a HVF. If
$\beta_{+}$ is set to zero in (\ref{szek}), then the resulting flat FRW model is in non-standard coordinates; although
there exists a HVF, finding it in this coordinate system is difficult due to the presence of the functions $a, b$ and
$c$. Determining a coordinate transformation into more familiar coordinates, where HVF's can easily be found, is also
difficult.  We use the following simplifying assumptions in (\ref{szek}), $a=1, b=0, c=0$, and consider deviations from
flat FRW by redefining $\beta_{+} \rightarrow \epsilon\beta_{+}$ for small $\epsilon$.  It can be shown that

\begin{equation}
\xi = \frac{\phi}{3}\left(T\frac{\partial}{\partial T}+x\frac{\partial}{\partial x}+y\frac{\partial}{\partial
y}+z\frac{\partial}{\partial z} \right)                                                            \label{hvfszek}
\end{equation}

\noindent is a HVF in the associated flat FRW model and deviates\footnote{As a special case, if $\beta_{+}=C/z^2$ then
(\ref{hvfszek}) is also a HVF of (\ref{szek}); therefore, as a consequence of \cite{hvfinv}, $\L_{\xi}P=0$.} from
homotheticity in (\ref{szek}) to first order in $\epsilon$.  Since $\lim_{T \rightarrow 0}(\L_{\xi}g_{ab}-2\phi
g_{ab})=0$ we find that (\ref{hvfszek}) becomes a HVF as the isotropic singularity is approached.  In this limit then
$\L_{\xi}P=0$, as expected \cite{hvfinv}, and so $P$ approaches a constant.  This type of behavior is desirable if we
would like to interpret $P$, in some sense, as a gravitational entropy possessing  a critical point at early times in
the evolution. Moreover, requiring that $P$ be monotonically increasing at early times along a timelike $\xi$
($\L_{\xi}P>0$) places restrictions on $\beta_{+}$.  Since

\begin{equation}
\L_{\xi}P=\frac{8}{9}\frac{\phi \epsilon^{2} T^4 \beta_{+}(2\beta_{+}+z\beta_{+}^{'})}{(1-\epsilon\beta_{+}
R^2)^3}=\frac{8}{9}\phi T^4 \beta_{+}(2\beta_{+}+z\beta_{+}^{'})\epsilon^{2}+O(\epsilon^3)   \label{liePxi}
\end{equation}
and
\begin{equation}
\xi\cdot\xi=\left(\frac{\phi}{3}\right)^{2}T^{4}\{-T^2 + \frac{R^2}{5}+[1-\epsilon\beta_{+}(R^2+T^2)]^2 z^2\}
\label{magxi}
\end{equation}

\noindent where $R^2 \equiv 5(x^2+y^2)$, then as $T \rightarrow 0$, $\xi$ remains timelike if $R=0$ and $z \rightarrow
0$ subject to the requirement that $\lim_{z\rightarrow 0}z\beta_{+}$ be bounded.  Assuming $z\beta_{+}$ is analytic
near $z=0$ gives the form

\begin{equation}
\beta_{+}=\frac{b_0}{z} + b_1 + b_2 z + \cdots     \label{bpser}
\end{equation}

\noindent To leading order in $\epsilon$ we factor $\beta_{+}$ from (\ref{liePxi}) and use (\ref{bpser}) to obtain
$\lim_{z \rightarrow 0}(2+z\beta_{+}^{'}/\beta_{+})=1$ thus $\L_{\xi}P >0$ along timelike $\xi$.

\section{Bianchi VI$_{h}$ model}
It has been shown \cite{wainander} for the Bianchi VI$_{h}$ class that a choice of parameters can result in the
quasi-isotropic stage beginning at the initial singularity, giving rise to an isotropic singularity for these
spacetimes. In the notation of \cite{wainander}, we set $\alpha_{s}=0$ and  $\alpha_{m}=1$, so that the line-element in
conformal time coordinates is

\begin{equation}
ds^2=\tau^{4/(3\gamma-2)}\left(-A^{2(\gamma-1)}d\tau^2+A^{2q_1}dx^2+ A^{2q_2}e^{2r[s+(3\gamma-2)]x}dy^2+
A^{2q_3}e^{2r[s-(3\gamma-2)]x}dz^2\right)    \label{bian6h}
\end{equation}

where

\begin{equation}
\begin{array}{cccccc}
A^{2-\gamma}=1+\alpha_{c}\tau^2,  & \displaystyle{q_{1}=\frac{\gamma}{2}}, & \displaystyle{q_{2}=\frac{2-\gamma+s}{4}},
& \displaystyle{q_{3}=\frac{2-\gamma-s}{4}}, & \displaystyle{s^2=(3\gamma+2)(2-\gamma)}, &
\displaystyle{r^2=\frac{(3\gamma+2)\alpha_{c}}{4(2-\gamma)(3\gamma-2)^2}}. \\
\end{array}
\end{equation}

These spacetimes have a perfect fluid source with equation of state $p=(\gamma-1)\mu$, $1\leq\gamma<2$.  The fluid
4-velocity is $u=A^{1-\gamma}\tau^{-2/(3\gamma-2)}\frac{\partial}{\partial\tau}$ and the energy density is
$\mu=12A^{-\gamma}(3\gamma-2)^{-2}\tau^{-6\gamma/(3\gamma-2)}$.  Since $\alpha_{s}$ has been set to zero, then the
isotropic singularity occurs at $\tau=0$.  The parameter $\alpha_{c}$ determines the curvature of the spacelike
hypersurfaces orthogonal to $u$, if $\alpha_{c}=0$ we obtain the flat FRW solution.  We shall consider deviations about
this flat FRW model by assuming $\alpha_{c}$ is small.

To leading order in $\alpha_{c}$, the gravitational epoch function for (\ref{bian6h}) is

\begin{equation}
P=\frac{4}{3}\frac{\gamma^2 (3\gamma-2)^2 \tau^4}{(\gamma-2)^2 [3(\gamma-1)^2+1]}\alpha_{c}^{2}+O(\alpha_{c}^{3}),
\label{bianP}
\end{equation}

\noindent which is positive and $P \rightarrow 0$ as $\tau \rightarrow 0^{+}$.  Using the Bel-Robinson tensor and the
energy density we find

\begin{equation}
\widetilde{P}=\frac{\gamma^2 (3\gamma-2)^2 \tau^4}{6(\gamma-2)^2}\alpha_{c}^{2}+O(\alpha_{c}^{3}).  \label{bianPbu}
\end{equation}

\noindent Consequently to leading order in $\alpha_c$ we have that $\widetilde{P}=[3(\gamma-1)^2+1]P/8$, and again $P$
and $\widetilde{P}$ are directly proportional. Unlike the Szekeres class above, the magnetic part of the Weyl tensor
with respect to $u$ does not vanish here unless\footnote{$H_{ab}$ will also vanish if $\alpha_{c}=0$.} $\gamma = 4/3$.
This relationship between $P$ and $\widetilde{P}$ becomes evident if we consider the expansions for the relevant
invariants of the electric and magnetic parts of Weyl

\begin{eqnarray}
E_{ab}E^{ab}=\frac{24\gamma^2}{(\gamma-2)^2 (3\gamma-2)^2 \tau^{8/(3\gamma-2)}}\alpha_{c}^{2}+O(\alpha_{c}^{3}) &,\ \
\displaystyle{H_{ab}H^{ab}=\frac{2\gamma^2 (3\gamma+2)(3\gamma-4)^2}{(2-\gamma)^3 (3\gamma-2)^2
\tau^{6(2-\gamma)/(3\gamma-2)}}\alpha_{c}^{3}+O(\alpha_{c}^{4})}.
\end{eqnarray}

\noindent For $\alpha_{c}$ small, $H_{ab}H^{ab}\sim 0$ and $C_{abcd}C^{abcd} \sim E_{ab}E^{ab} \sim W$, additionally
Einstein's equations give $R_{ab}R^{ab}=[3(\gamma-1)^2+1]\mu^2$, the relationship now follows.

As $\alpha_{c} \rightarrow 0$ or if $\tau \rightarrow 0^+$, the vector

\begin{equation}
\xi = \frac{\phi}{3}\frac{(3\gamma-2)}{\gamma}\left(\tau\frac{\partial}{\partial\tau}+x\frac{\partial}{\partial
x}+y\frac{\partial}{\partial y}+z\frac{\partial}{\partial z} \right)                        \label{hvfbian6}
\end{equation}

\noindent gives $(\L_{\xi}g_{ab}-2\phi g_{ab})\rightarrow 0$ separately in both limits; therefore, $\xi$ is a HVF in
the flat FRW limit or as the isotropic singularity is approached.  Setting $x=y=z=0$ we obtain for the magnitude of
$\xi$

\begin{equation}
\xi\cdot\xi=-\left(\frac{\phi}{3}\right)^{2}\frac{(3\gamma-2)^2}{\gamma^2}\tau^{6\gamma/(3\gamma-2)}A^{2(\gamma-1)};
\end{equation}

\noindent thus by continuity $\xi$ will be timelike in some neighborhood of $x=y=z=0$ arbitrarily close to the
isotropic singularity.  To leading order in $\alpha_c$, the behavior of $P$ along $\xi$ is given by

\begin{equation}
\L_{\xi}P=\frac{16}{9}\frac{\phi\gamma(3\gamma-2)^3 \tau^4}{(\gamma-2)^2
[3(\gamma-1)^2+1]}\alpha_{c}^{2}+O(\alpha_{c}^{3})
\end{equation}

\noindent which is positive for $1\leq\gamma<2$; hence $P$ will be monotonically increasing at early times along the
timelike $\xi$, and as $\tau \rightarrow 0^+$ then $P$ will tend to a constant (which in these models is zero).

We now show that the Weyl curvature hypothesis does not necessarily put restrictions on the Petrov type.  In this class
of spacetimes the Weyl invariants of CM have vanishing imaginary parts, and in general do not always satisfy the syzygy
$w_{1}^{3}-6w_{2}^{2}=0$; therefore, the metric(\ref{bian6h}) is almost always Petrov type I.  However for
$4/3<\gamma<2$ there always exists a time $\tau_{*}$ where the Petrov type specializes to either II or D; this is given
by

\begin{equation}
\tau_{*}=\frac{3}{2}\frac{2-\gamma}{\sqrt{2\alpha_{c}(3\gamma-4)}},
\end{equation}

\noindent otherwise the Petrov type is I.  When $\gamma=4/3$ the syzygy is satisfied and by choosing the aligned
Newman-Penrose tetrad
\begin{equation}
\begin{array}{ccc}
\displaystyle{\ell=A^{-2/3}\frac{\partial}{\partial\tau}+\frac{\partial}{\partial z}}, &
\displaystyle{n=\frac{1}{2t^2}\left(\frac{\partial}{\partial\tau}-A^{2/3}\frac{\partial}{\partial z} \right)}, &
\displaystyle{m=\frac{1}{\sqrt{2}\tau A^{2/3}}\left(\frac{\partial}{\partial x}+i e^{-4rx}\frac{\partial}{\partial y}
\right)}
\end{array}
\end{equation}

\noindent we find that this is in fact Petrov type D for all $\tau>0$.  It would appear that an intermediate algebraic
specialization of the Weyl tensor during the evolution does not affect the increasing anisotropy that is indicated by
the gravitational epoch function (\ref{bianP}); indeed, these spacetimes begin with small anisotropy close to the
isotropic singularity and approach the anisotropic vacuum plane wave metrics at late times.

As is shown in \cite{hvfinv}, \emph{any} dimensionless ratio of invariants is constant along a HVF. Therefore,
depending on its monotonicity, it may also serve as a gravitational epoch function when the cosmological model admits
an asymptotic timelike HVF.  Here we illustrate this point by considering a dimensionless ratio of differential
invariants; to second order in $\alpha_c$ we have

\begin{equation}
P_{1}=\frac{\nabla_{a}C_{bcde}\nabla^{a}C^{bcde}}{\nabla_{a}R_{bc}\nabla^{a}R^{bc}}=\frac{40}{9}\frac{(3\gamma-2)^{2}
\tau^4} {(\gamma-2)^2 [9(\gamma-1)^2 +5]}\alpha_{c}^2 + O(\alpha_{c}^{3})        \label{bianP1}
\end{equation}

and

\begin{equation}
\L_{\xi}P_{1}=\frac{160}{27}\frac{\phi(3\gamma-2)^3 \tau^4}{\gamma(\gamma-2)^2 [9(\gamma-1)^2 +5]}\alpha_{c}^{2} +
O(\alpha_{c}^{3}).                                                      \label{bianLP1}
\end{equation}

\noindent Clearly as $\tau \rightarrow 0^{+}$, $P_{1} \rightarrow 0$ and from (\ref{bianLP1}) it is also monotonically
increasing for $1\leq\gamma <2$.  Nevertheless, the invariants of (\ref{bianP1}) diverge in a similar manner to the
invariants of (\ref{bianP}), i.e. as $\tau \rightarrow 0^{+}$,  $\nabla_{a}C_{bcde}\nabla^{a}C^{bcde}$ and
$\nabla_{a}R_{bc}\nabla^{a}R^{bc}$ $\rightarrow -\infty$.

\section{Discussion}
We have considered a class of spatially inhomogeneous Szekeres solutions with the line-element (\ref{szek}). We show
that there exists a HVF (\ref{hvfszek}) as the isotropic singularity is approached, and show that  $\L_{\xi}P=0$ in
this limit, so that $P$ approaches a constant as expected \cite{hvfinv}. Moreover, to leading order we show that
$\L_{\xi}P
>0$ along an approach to the singularity in which the HVF $\xi$ remains timelike; i.e.,  $P$ is monotonically
increasing at early times along $\xi$. We then considered a class of Bianchi VI$_{h}$ models with an isotropic
singularity with line-element (\ref{bian6h}), parameterized by $\alpha_{c}$ (which measures deviations about the flat
FRW model). Assuming $\alpha_{c}$ is small, to leading order we display a vector $\xi$ which is a HVF in the flat FRW
limit, and show that the gravitational epoch function $P \rightarrow 0$  as the isotropic singularity is approached and
that $P$ is monotonically increasing at early times along timelike $\xi$.

Therefore, in the isotropic singularity cosmological models we have studied we have found that $ \textit{P}\rightarrow
0$ asymptotically as the self-similar cosmological model is approached, in support of the idea that these
homothetically self-similar spacetimes represent asymptotic equilibrium states. Moreover, we have provided evidence
that $P$ is monotonically increasing as the models evolve away from these equilibrium states, which perhaps lends
support to the idea that  $P$  represents a "gravitational entropy".

We also found that for both the Szekeres models and the Bianchi VI$_{h}$ models (to leading order in $\alpha_{c}$), the
standard gravitational epoch function $P$, and the normalized Bel-Robinson epoch function $\widetilde{P}$, are
proportional (and hence equivalent as gravitational epoch functions).  The question remains as to whether this will be
true for all models with an isotropic singularity.  In general, for a perfect fluid source we have that
$\widetilde{P}\sim (E^2 + H^2)/\mu^2$ and $P \sim (E^2-H^2)/(\mu^2 + 3p^2)$.  Assuming an equation of state of the form
$p=\alpha\mu$ gives $P\sim (E^2-H^2)/[(1+3\alpha^2 )\mu^2]$.  Clearly if $H^2$ is negligible with respect to $E^2$,
then $P$ and $\widetilde{P}$ will be effectively proportional.  It may also be of interest to consider cosmological
models where $P$ and $\widetilde{P}$ differ.  In particular, whenever the Petrov type is III, N or O then all zeroth
order Weyl invariants vanish, and hence $P$ vanishes but $\widetilde{P}$ does not necessarily vanish. An example of
such cosmological models are the Oleson \cite{exsol} solutions, which are Petrov type N with a perfect fluid source. In
these models $P$ vanishes but $\widetilde{P}$ does not; it is of interest to determine if these models can admit an
isotropic singularity.

A classic problem in cosmology is finding a way to explain the very high degree of isotropy observed in the cosmic
microwave background.  In GR cosmological models admitting an isotropic singularity are of zero measure, so that
isotropy is a special rather than generic feature of cosmological models. Hence, a dynamical mechanism which is able to
produce isotropy, such as inflation, is needed. However, inflation requires sufficiently homogeneous initial data in
order to begin \cite{KT}; hence the isotropy problem remains open to debate in standard cosmology. Recently, it has
been argued that an isotropic singularity is typical in brane world cosmological models \cite{coley}. Hence brane
cosmology would have the very attractive feature that it provides for the necessary sufficiently smooth initial
conditions which might, in turn, be consistent with entropy arguments and the second law of thermodynamics.

\section{Acknowledgements}
We would like to thank K. Lake for helpful discussions.  AAC is supported by NSERC of Canada.

\end{document}